# Pulse-width and Temperature Dependence of Memristive Spin-Orbit Torque Switching

Wei-Bang Liao[1], Tian-Yue Chen[1], Yu-Chan Hsiao[1], and Chi-Feng Pai[1,2] *

[1] Department of Materials Science and Engineering, National Taiwan University, Taipei 10617, Taiwan

[2] Center of Atomic Initiative for New Materials, National Taiwan University, Taipei 10617, Taiwan

It is crucial that magnetic memory devices formed from magnetic heterostructures possess sizable spin-orbit torque (SOT) efficiency and high thermal stability to realize both efficient SOT control and robust storage of such memory devices. However, most previous studies on various types of magnetic heterostructures have focused on only their SOT efficiencies, whereas the thermal stabilities therein have been largely ignored. In this work, we study the temperature-dependent SOT and stability properties of two types of W-based heterostructures, namely W/CoFeB/MgO (standard) and CoFeB/W/CoFeB/MgO (field-free), from 25 °C (298 K) to 80 °C (353 K). Via temperature-dependent SOT characterization, the SOT efficacies for both systems are found to be invariant within the range of studied temperatures. Temperature-dependent current-induced SOT switching measurements further show that the critical switching current densities decrease with respect to the ambient temperature; thermal stability factors ($\Delta$) are also found to degrade as temperature increases for both standard and field-free systems. The memristive SOT switching behaviors in both systems with various pulse-widths and temperatures are also examined. Our results suggest that although the SOT efficacy is robust against thermal effects, the reduction of $\Delta$ at elevated temperatures could be detrimental to standard memory as well as neuromorphic (memristive) device applications.

* Email: cfpai@ntu.edu.tw

In recent years, due to the fast development of both complementary metal-oxide-semiconductor (CMOS) and magnetic technologies, the realization of magnetic random access memory (MRAM) has gained lots of interests. Not only with the benefits such as low power consumption, short write time, non-volatility, and high endurance, MRAM can also potentially replace contemporary memory technologies. For standard memory applications, there are two types of MRAMs with different writing modes: spin transfer torque MRAM (STT-MRAM)[1] and spin-orbit torque MRAM (SOT-MRAM)[2]. The latter is more favorable than the former due to the minimal degradation of the tunnel barrier of its writing principle. The mechanism behind SOT-MRAM to control the magnetization of its ferromagnetic layer (FM) is the bulk spin Hall effect (SHE)[3-5] and/or the interfacial Rashba effect[6-9]. Either origin, the strong spin-orbit coupling of the heavy metal (HM) or at the HM/FM interface when a longitudinal current is applied to a magnetic heterostructure will generate a transverse spin current. The spin current flowing towards the adjacent FM will then exert a spin-torque upon it, and then switch the magnetization therein. Therefore, typically a SOT-MRAM layer stack consists of HM/FM/MgO/FM multilayers, where the applied charge current for writing will only flow in the HM layer. In contrast, the writing current needs to flow through the MgO tunnel barrier for the STT case to achieve FM moment switching. However, it is important to note that for heterostructures or devices with FM having perpendicular magnetic anisotropy (PMA), an external in-plane magnetic field is required to break the symmetry to achieve deterministic current-induced SOT switching.[5,6] To realize the "field-free" SOT switching, some specific mechanism must be adopted and additional engineering on the layer design is necessary, such as introducing exchange bias,[10,11] wedged structure,[12-16] interlayer exchange coupling,[17,18] or Néel orange-peel effect.[19,20]

Presently many of the SOT-related studies focus on enhancing SOT efficiency and realizing deterministic field-free SOT switching through layer stack engineering. However, magnetic heterostructures or devices with good thermal stability is also a key factor to make such memory applications useful. Especially when these devices are integrated with CMOS, the thermal dissipation from transistors may increase the device temperature and affect its SOT efficiency and thermal stability factor ($\Delta$), which needs to be scrutinized. It is known that the probability of a magnetic state that can be flipped due to thermal fluctuation is proportional to $\exp(-\Delta/k_B T)$.[21] However, most previous works focused on the $\Delta$ at 25 °C (298 K) without considering the temperature effect. On the other hand, SOT-based devices utilized in neuromorphic applications have also drawn lots of attention due to their potential in achieving memristive switching.[22-24] Previous works have shown that the memristive switching can be tuned by the amplitude and the duration of the applied current at room temperature.[23,25] However, the memristive switching behavior at elevated temperatures has yet to be carefully examined.

In this work, we systematically characterize the SOT efficacy $\chi$ (in terms of effective field per current density) and the thermal stability factor $\Delta$ from $T$ = 25 °C (298 K) to 80 °C (353 K) of a typical Hall bar structure W/CoFeB/MgO (standard) and a robust field-free switching Hall bar structure CoFeB/W/CoFeB/MgO (field-free). We find that $\chi$ of both systems remain fairly constant at elevated temperatures. Current-induced SOT switching measurements are performed to characterize $\Delta$ via two approaches: pulse-width-dependent and temperature-dependent measurements.[26] As expected, the critical switching current density reduces with increasing ambient temperature or applied current pulse-width. $\Delta$ also gradually decreases as temperature rises to 80 °C (353 K). $\Delta$ extracted from the two approaches are comparable. Furthermore, we study the memristive switching behavior of these devices with various durations and amplitudes of current pulses at different temperatures. The *limits* of memristive switching window, as defined by the maximum and the minimum current densities to obtain intermediate states, can be adjusted by the applied current pulse-width. The *size* of the memristive window is largely unaffected by changing current pulse-width at 25 °C (298 K), however, a significant reduction will take place at 80 °C (353 K). This trend of memristive switching behavior is also observed in the field-free device. Our temperature-dependent results pose the potential issues of using SOT devices, either standard or field-free, for memristive or neuromorphic applications at elevated temperatures.

In this work, we prepare two kinds of samples to characterize their temperature-dependent SOT efficiencies and current-induced switching behaviors at elevated temperatures. One is the standard sample, W(4)/CoFeB(1.4)/MgO(2)/Ta(2), and the other being the field-free switching sample CoFeB(3)/W(1.5)/CoFeB(1.4)/MgO(1.6)/Ta(2) (units in nanometers). Both samples are deposited in multilayer stacks on amorphous Si/SiO$_2$ substrates via high vacuum magnetron sputtering with a base pressure of $3\times10^{-8}$ Torr. DC and RF magnetron sputtering with 3 mTorr and 10 mTorr of Ar working pressure are employed for depositing metallic and oxide layers, respectively. The standard sample is in-situ deposited at 300 °C without further heat treatments while the field-free switching sample is deposited at 25 °C and then post-annealed at 300 °C for 1 hr. Both samples show PMA.

In order to characterize the SOT-induced effective fields at elevated temperatures, we perform current-induced hysteresis loop shift measurements[27] on micron-sized Hall bar devices patterned via standard photolithography followed by Ar ion-mill etching. The width ($w$) of the Hall bar device is 10 μm for the standard sample and 5 μm for the field-free case. As schematically shown in Fig. 1(a), we sweep the out-of-plane magnetic field and apply a charge current (density) $J_{dc}$ along x-direction. The in-plane field $H_x$ is applied to overcome the Dzyaloshinskii-Moriya interaction (DMI) effective field $H_{DMI}$, such that the Néel chiral domain wall moments in the magnetic layer can be realigned. Then, the spin current generated from the SHE of the W(4) layer transfers spin-torque onto the realigned domain wall moments and brings about the domain wall propagation and domain expansion.[28,29] As shown in Fig. 1(b), the current-induced effective field $H_z^{eff}$ causes the out-of-plane hysteresis loop to shift along $H_z$ in a direction that depends on the applied current polarity. When $H_x \geq H_{DMI}$, the SOT generated from W can fully act on the adjacent CoFeB moments, where the figure of merit $\chi \equiv H_z^{eff} / J_{dc}$ (effective field per current density) reaches a saturated value, as shown in Fig. 1(c). $J_{dc}$ in this work represents the current density flowing in the W layer. As estimated from the resistivities of W ($\rho_W \approx 180$ μΩ-cm) and CoFeB ($\rho_{CoFeB} \approx 200$ μΩ-cm) layers, the current flowing in the W layer should be ~ 76 % of the total applied current for the standard sample and ~ 27 % for the field-free sample. The maximum magnitude of $|\chi| \approx 63$ Oe/$10^{11}$ A·m$^{-2}$ for the standard sample and 21 Oe/$10^{11}$ A·m$^{-2}$ for the field-free switching sample are consistent with previous reports.[30,31] Moreover, for the field-free switching device, we perform measurements with different pre-magnetizing fields, *i.e.*, the in-plane field is swept with different sequences: (1) $H_x = -1500$ Oe $\rightarrow 0 \rightarrow 1500$ Oe, and (2) $H_x = 1500$ Oe $\rightarrow 0 \rightarrow -1500$ Oe. It can be clearly seen that there are non-zero SOT efficacies $\chi \approx \pm 7$ Oe/$10^{11}$ A·m$^{-2}$ at $H_x = 0$ and the sign is opposite for these two sequences as shown in the inset of Fig. 1(c), which suggests that the symmetry breaking effective field is parallel to the bottom in-plane CoFeB(3) magnetization direction. In addition, we perform current-induced SOT switching measurement on field-free switching sample. By pre-magnetizing the sample with a large positive or negative in-plane external field ($\pm 1000$ Oe), the polarity of these two case are opposite (data not shown here). This further confirms the existence of the symmetry-breaking effective field, and the mechanism of achieving field-free SOT switching can be attributed to the interlayer exchange coupling[17,18,32] or the Néel orange-peel effect[19,20] between two FMs.

Then we use a custom-made heater to heat up these devices and perform hysteresis loop shift measurements from 30 °C (303 K) to 70 °C (343 K). $\chi$ of the standard sample is characterized with applying an in-plane field $H_x = 300$ Oe while no $H_x$ is applied for the field-free switching sample. We find that $\chi$ is independent of temperature for both samples, which indicates that the SOT efficacy of W is invariant for temperature up to 70 °C (343 K), as shown in Fig. 1(d). Note that the calculated damping-like SOT efficiency ($\xi_{DL} \propto M_s t_{CoFeB} \chi$) or the spin Hall ratio would only slightly decrease due to the degradation of saturation magnetization ($M_s$) while raising temperature. Similar results have also been demonstrated for the Pt case.[33] However, there exist contrary results indicating that the SOT efficiency of Ta heterostructures will significantly reduce at $T > 300$ K.[34] Nevertheless, our results suggest that the temperature effect should have a minimal influence on the SOT efficacy or efficiency of both standard and field-free W-based SOT devices.

Next, to study the thermal stabilities of these devices for memory applications, we perform current-induced SOT switching measurements at various current pulse-widths and temperatures. Fig. 2(a) shows a schematic diagram of a patterned standard device with current pulse injected into the 10-μm-wide channel, where the magnetization switching can be monitored by measuring the change of anomalous Hall voltage $(V_H)$. A small in-plane field $H_x = 80$ Oe is applied to overcome $H_{DMI}$ and to ensure deterministic switching. Representative current-induced switching results with different pulse-widths and temperatures are displayed in Fig. 2(b), (c), respectively. It can be observed that the switching current density ($J_c$) monotonically decreases with either increasing pulse-width (Fig. 2(b)) or increasing temperature (Fig. 2(c)). Typically the spin torque switching behavior can be divided into a dynamical and a thermally-activated regime, depending on the pulse-width applied.[35,36] For our case (50 μs $\leq t_{pulse} \leq$ 100 ms at 80 °C (353 K)), the switching is within the thermally-activated regime and the critical switching current density is expressed as[37]

$$J_c = J_{c0}\left[1 - \frac{1}{\Delta}\ln\left(\frac{t_{\text{pulse}}}{\tau_0}\right)\right], \quad (1)$$

where $J_{c0}$ is the zero thermal critical switching current density, $\Delta \equiv U/k_B T$ is the above-mentioned thermal stability factor, $U$ is the energy barrier between two magnetization states,[38] and $1/\tau_0$ ($\tau_0 \approx 1$ ns) is the intrinsic attempt frequency.[36] Fig. 2(d) shows the linear fits of experimental data at different temperatures for $J_c$ versus $\ln(t_{\text{pulse}}/\tau_0)$. Note that since the thermal effect affects the range of thermally-activated switching regime, the linear fitting range (where $R^2 > 0.9$ is satisfied) depends on the ambient temperature. For the micron-sized device, the shortest pulse-width in fitting range is tens of μs, which is longer than works focused on nano-sized device.[24,39] Fig. 2(e) displays the temperature dependence of $J_c$ from 25 °C (298 K) to 90 °C (363 K) with a constant pulse-width of $t_{\text{pulse}} = 5$ ms. Again, by performing the linear fitting of the experimental data with eqn. (1), the extracted energy barrier $U \approx 0.56$ eV, which corresponds to $\Delta \approx 22$ at 25 °C (298 K). Also note that the temperature change due to Joule heating from the applied pulse current is less than 2 K, as characterized by temperature-dependent and current-dependent resistance measurements.[40] Therefore, the current-induced heating effect is treated as a negligible factor in this work. However, even though the Joule heating is not significant in our micron-sized Hall bar devices, it could be a more pronounced effect as the device becomes nano-sized.[41]

We summarize $\Delta$ and $J_{c0}$ obtained from both pulse-width and temperature-dependent switching measurements in Fig. 2(f). Several remarks can be made from these results: (1) Since the experimental data of Fig. 2(e) (changing temperature with pulse-width fixed) is well linearly-fitted, $U$ should be a constant within the studied temperature range, which is quite consistent with the results by Rahaman et al.[26] Therefore, $\Delta$ should monotonically decrease with increasing temperature, which is denoted as the dash line shown in Fig. 2(f). (2) From pulse-width-dependent measurements, $\Delta$ extracted at 25 (298), 50 (323), and 80 (353) °C (K) gradually decreases, which are comparable to the $\Delta$ obtained from temperature-dependent-method. (3) The calculated spin-torque switching efficiency[42] $\varepsilon \equiv \Delta/J_{c0}$ reduces at elevated temperatures. (4) As shown in Fig. 2(f), $|J_{c0}|$ under 25 (298), 50 (323), and 80 (353) °C (K) are all around $9.3 \times 10^{10}$ A/m$^2$ and show little change within the tested temperature range. This is consistent with the fact that $J_{c0}$ corresponds to the overall spin angular momentum needed to overcome damping and energy barrier excluding thermal effects.[43] Hence, $J_{c0}$ should be independent of the device temperature, which is also in line with our discovery of $\chi$ being temperature independent.

Next, the memristive switching behavior[22] of the standard sample is investigated. Before applying pulsed current, the magnetization is reset to the magnetization-up state via an out-of-plane field which is larger than the coercivity. This is followed by sweeping the pulse current $-J_{pulse}^{max} \rightarrow 0 \rightarrow J_{pulse}^{max} \rightarrow 0 \rightarrow -J_{pulse}^{max}$. The cycle is then repeated for regulating $J_{pulse}^{max}$ until full magnetization switching is achieved. Fig. 3(a) shows some representative $V_H$-$J_{pulse}$ hysteresis loops with $t_{pulse} = 5$ μs at room temperature, in which the $J_{pulse}^{max}$-dependent intermediate states of $V_H$ are observed. Fig. 3(b) shows the applied current pulse-width dependence of memristive switching behavior. With reducing pulse-width, the amplitude of current needed to promote the formation of magnetic domains and domain wall propagation for magnetization switching is enhanced. Take the $t_{pulse} = 5$ μs and $t_{pulse} = 500$ μs data in Fig. 3(b) for examples, the memristive switching window is $2.6 \times 10^{10}$ A/m$^2 \leq J_{pulse}^{max} \leq 5.4 \times 10^{10}$ A/m$^2$ ($\Delta J_{pulse}^{window} \approx 2.8 \times 10^{10}$ A/m$^2$) for $t_{pulse} = 5$ μs and is $1.0\times10^{10}$ A/m$^2$ $\leq J_{\text{pulse}}^{\max} \leq 3.5\times10^{10}$ A/m$^2$ ($\Delta J_{\text{pulse}}^{\text{window}} \approx 2.5\times10^{10}$ A/m$^2$) for $t_{\text{pulse}} = 500$ µs, which suggests that the upper and lower limits of the memristive switching window depends on the applied current pulse-width, yet the window size of obtaining intermediate states ($\Delta J_{\text{pulse}}^{\text{window}}$) remains fairly the same. To further verify the repeatability of achieving intermediate states, we apply three different amplitudes of pulsed currents ($J_{\text{pulse}} = 3.04\times10^{10}$, $2.66\times10^{10}$, and $1.90\times10^{10}$ A/m$^2$) for ~ 20 cycles, as shown in Fig. 3(c). The final states are fairly stable and correspond to the circled intermediate states in Fig. 3(b), which suggests that achieving these memristive intermediate states is repeatable. Note that most of the reported memristive switching behavior, including the results present here, are mainly governed by domain wall motion in micron-sized devices, which have also been observed in the conventional Ta/CoFeB/MgO and Pt/Co PMA systems.[23,25] However, it is probable that as the device size downscales to several nanometers and becomes single domain, such memristive switching behavior will become more discretized and therefore imposes challenges in neuromorphic applications.

Fig. 3(d) shows representative memristive switching behavior of the standard sample at different temperatures. The rising temperature promotes domain nucleation and domain wall propagation via enhanced thermal fluctuation, therefore $J_c$ decreases as increasing temperature. The memristive switching window shifts towards the lower amplitude of pulsed current. Also, the saturation magnetization slightly reduces at 80 °C (353 K), which can be observed in Fig. 3(d) that the maximum switching percentage reaches $\approx 80$ % of the original room-temperature $\Delta V_H$. More importantly, with increasing temperature and more pronounced thermal fluctuations, $\Delta J_{\text{pulse}}^{\text{window}}$ decreases from $\approx 2.8\times10^{10}$ A/m$^2$ at 25 °C (298 K) to $\approx 1.5\times10^{10}$ A/m$^2$ at 80 °C (353 K). This result suggests that the temperature range at which typical micro-processors operate (~ 40 °C to 90 °C) might already undermine the feasibility of employing such SOT devices for memristive applications.

  We now turn focus to the field-free switching sample. A schematic diagram of such Hall bar device is shown in Fig. 4(a). No in-plane magnetic field is applied (using an electromagnet-free probe station) for both pulse-width and temperature-dependent measurements to characterize its thermal stability and memristive switching behavior. Again according to eqn. (1), by performing linear fits to the experimental data in Fig. 4 (b) with various pulse-widths at 25 °C (298 K), $\Delta \approx 39$ and $J_{c0} \approx 1.14 \times 10^{11}$ A/m$^2$ of this field-free device are extracted. As shown in Fig. 4(c), from the temperature-dependent results with constant pulse-width $t_{\text{pulse}} = 50$ ms, we extract $J_{c0} \approx 1.57 \times 10^{11}$ A/m$^2$ and $U \approx 0.76$ eV, which translates to $\Delta \approx 30$ at 25 °C (298 K). Therefore, $\Delta$ and $J_{c0}$ obtained from these two different methods are comparable in our field-free switching sample, same as the standard sample case and previous studies.[26] It is worth noting that the field-free switching sample has a higher room-temperature memory retention ($\Delta \sim 30$) compared to standard sample ($\Delta \sim 20$), which is possibly related to the higher coercivity (18 Oe vs. 11 Oe).

  Finally, we study the field-free memristive switching behavior of the field-free sample. Fig. 4(d) shows representative $V_H$ - $J_{\text{pulse}}$ hysteresis loops with $t_{\text{pulse}} = 50$ ms measured at zero field and 25 °C. Similar to the standard case, different $J_{\text{pulse}}^{\max}$ leads to different intermediate states. The pulse-width-dependent field-free memristive switching behavior at 25 °C (298 K) is displayed in Fig. 4(e). It can be seen that the size of memristive switching window does not show a great change for $t_{\text{pulse}} = 50, 100, 500$ ms, and the numbers of accessible intermediate states are almost the same. Therefore, stable memristive applications could be performed using tens to hundreds ms pulse-width currents in this field-free switching device, at least at room temperature. Fig. 4(f) shows the memristive switching behavior from 25 °C (298 K) to 80 °C (353 K). As in the standard sample case, the number of intermediate states reduces with rising temperature. $\Delta J_{\text{pulse}}^{\text{window}}$ decreases from $\approx 5\times10^{10}$ A/m$^2$ at 25 °C (298 K) to $\approx 2\times10^{10}$ A/m$^2$ at 80 °C (353 K), which again indicates a lower applicability of SOT memristive applications at elevated temperatures.

In summary, we systematically analyze the SOT efficacy and switching behavior of two types of W-based heterostructures at different temperature. We observe that the SOT efficacies (in terms of effective field per current density $\chi$) remain invariant from 30 °C (303 K) to 70 °C (343 K) for both standard (W/CoFeB/MgO) and field-free (CoFeB/W/CoFeB/MgO) magnetic heterostructures. Then through current-induced SOT switching measurements, we show critical switching current density decreases with increasing current pulse-width or raising temperature. We further analyze $\Delta$ and $J_{c0}$ via pulse-width and temperature-dependent measurements, where $\Delta$ is found to decrease as increasing temperature while $J_{c0}$ remains constant. Moreover, the results obtained from these two methods are fairly consistent. We also demonstrate the memristive switching behavior from both standard and field-free switching devices. The memristive switching window of applied current to achieve intermediate states can be affected by current pulse-width and device temperature. More importantly, for both standard and field-free structures, the size of the memristive widnow will shrink as $T$ increases. This will pose challenges to potential memristive or neuromorphic applications using SOT devices at elevated temperatures.

**Acknowledgements**

This work was supported by the Ministry of Science and Technology of Taiwan (MOST) under grant No. 109-2636-M-002-006, and by the Center of Atomic Initiative for New Materials (AI-Mat), National Taiwan University from the Featured Areas Research Center Program within the framework of the Higher Education Sprout Project by the Ministry of Education (MOE) in Taiwan under grant No. NTU-107L9008.

## Data Availability

The data that support the findings of this study are available from the corresponding author upon reasonable request.

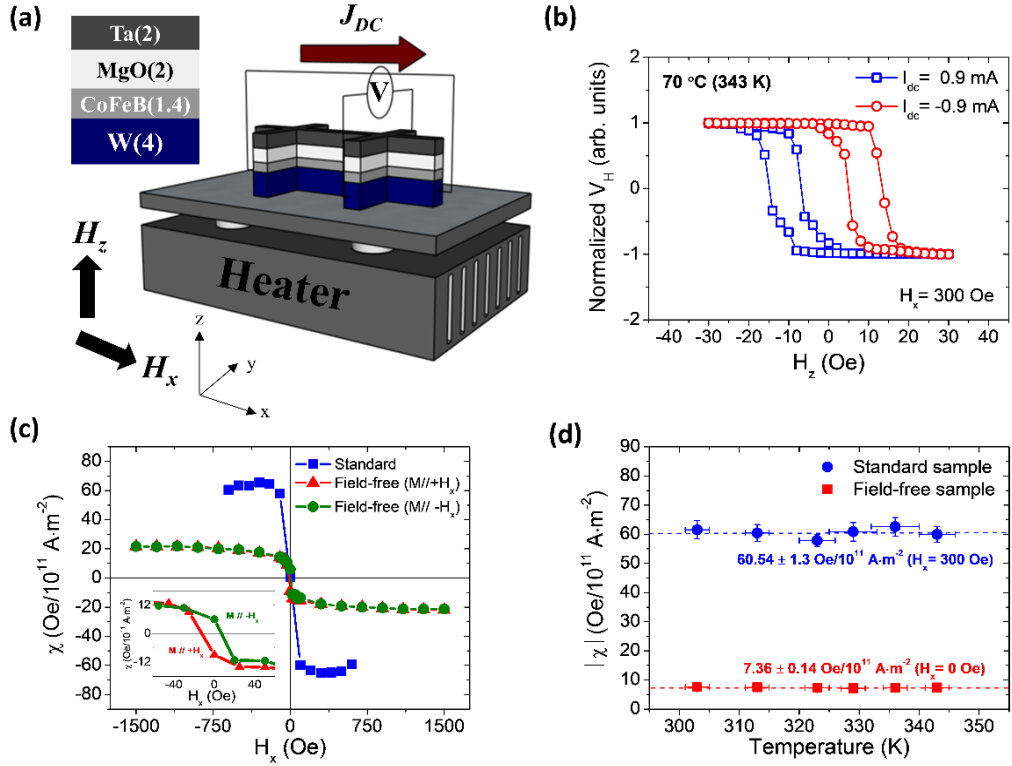

FIG. 1. (a) Schematic illustration of a standard Hall bar device (with layer structure: W(4)/CoFeB(1.4)/MgO(2)/Ta(2)) for hysteresis loop shift measurement. (b) Representative hysteresis loop shifts of a standard device. (c) $\chi$ (SOT efficacy) as a function of $H_x$ for standard sample (W(4)/CoFeB(1.4)/MgO(2)/Ta(2), blue square) and field-free switching sample (CoFeB(3)/W(1.5)/CoFeB(1.4)/MgO(1.6)/Ta(2), green circle and red triangle). The red triangle (green circle) represents the data with in-plane CoFeB(3) layer premagnetized along $+H_x$ ($-H_x$). The inset shows $|H_x| \leq 50$ Oe data of the field-free switching sample with two different pre-magnetizing directions. (d) $|\chi|$ as a function of temperature of both samples.

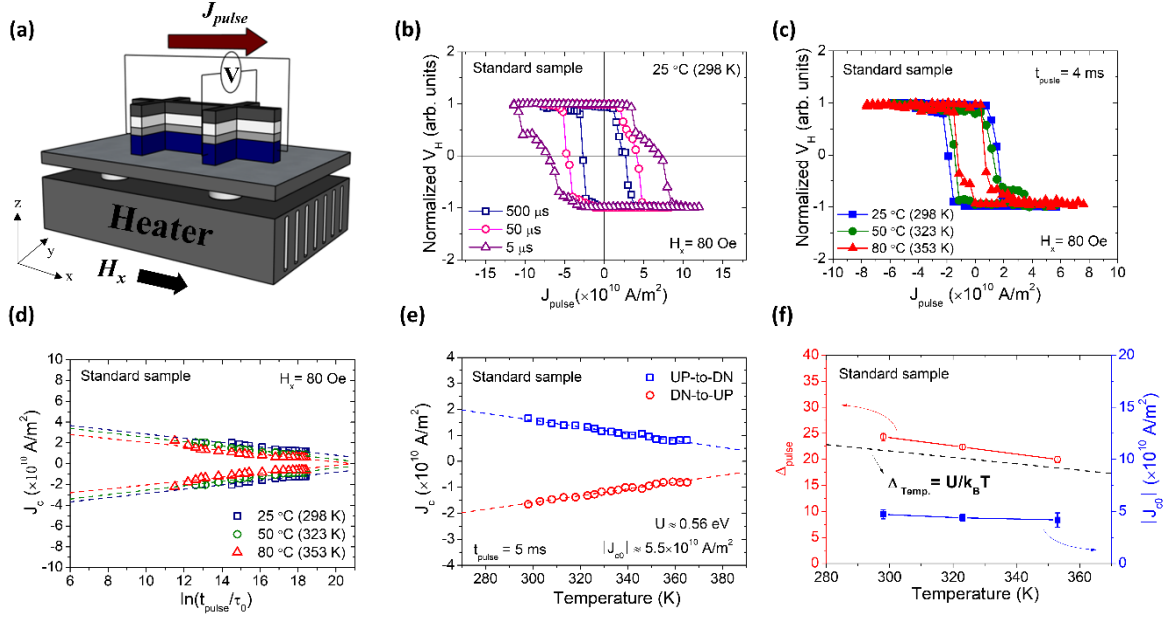

FIG. 2. (a) Schematic illustration of current-induced SOT switching measurement on a standard sample Hall bar device (W(4)/CoFeB(1.4)/MgO(2)/Ta(2)). Representative current-induced switching loops (b) with pulse-width dependence at 25 °C (298 K) and (c) with temperature dependence at $t_{\text{pulse}} = 4$ ms. Critical switching current densities as functions of (d) pulse-width at 25 (298), 50 (323), and 80 (353) °C (K) and (e) temperature with $t_{\text{pulse}} = 5$ ms. The dash lines represent linear fits of the experimental data. (f) Thermal stability factor and zero thermal critical switching current density obtained from pulse-width dependent measurements as a function of temperature. The black dash line represents the thermal stability factor as a function of temperature obtained from temperature dependent measurements.

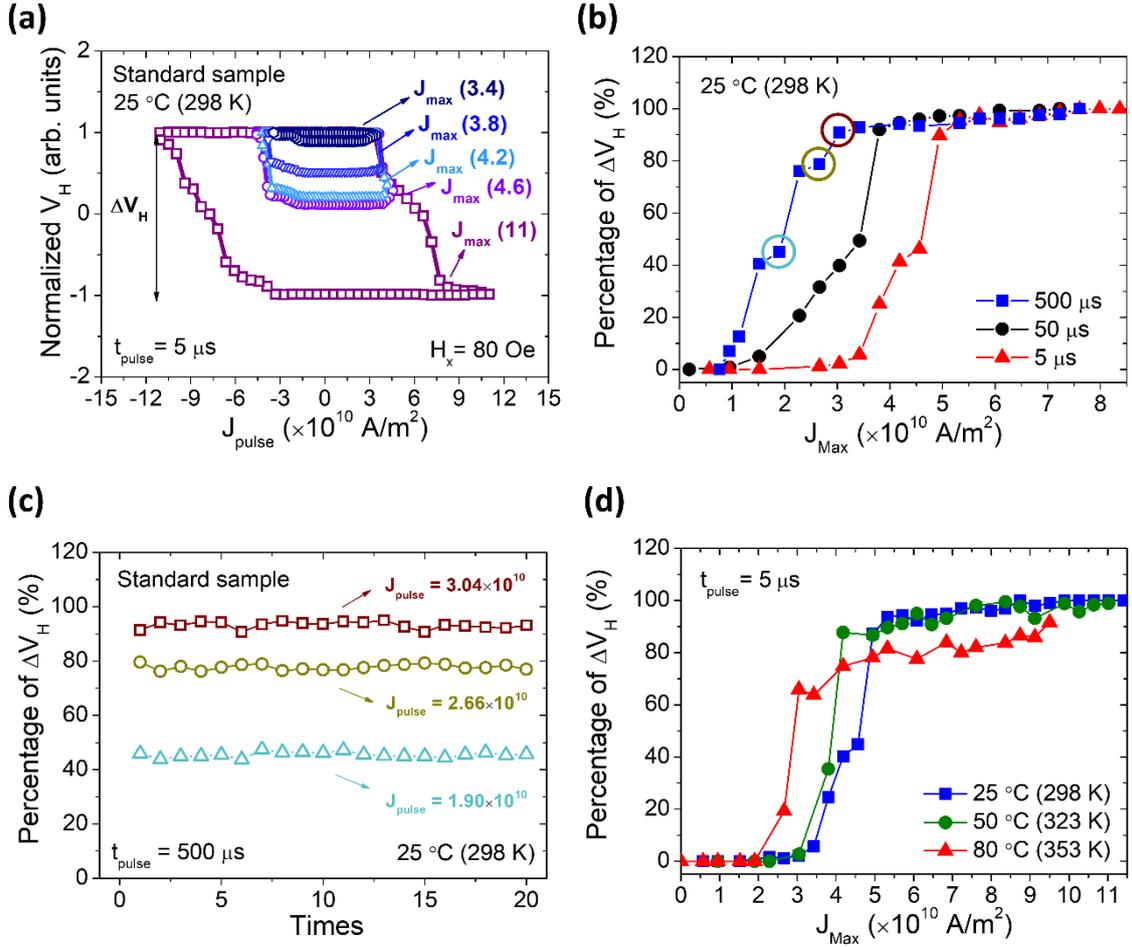

FIG. 3. (a) The memristive current-induced switching behavior of a standard device with various $J_{max}$ (in units of $10^{10}$ A/m$^2$). (b) The switching percentages in terms of anomalous Hall voltage as functions of the applied $J_{max}$ for pulse-width dependent measurements. (c) The repeatability of achieving three different intermediate states corresponding to the circled values in (b) with constant amplitude of pulsed currents. (d) The switching percentages in terms of anomalous Hall voltage as functions of the applied $J_{max}$ for temperature dependent measurements

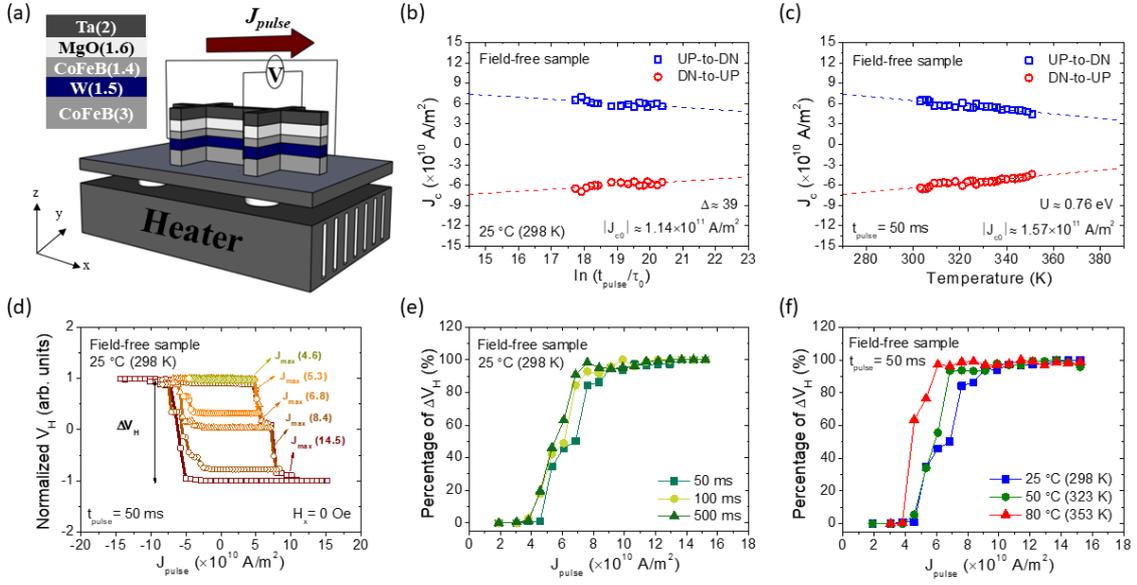

FIG. 4. (a) Schematic illustration of current-induced SOT switching measurements on a field-free switching device (CoFeB(3)/W(1.5)/CoFeB(1.4)/MgO(1.6)/Ta(2)) with no in-plane field applied. Switching current densities (b) with various pulse-widths at 25 °C (298 K) and (c) various temperatures at $t_{pulse} = 50$ ms. The dash lines represent linear fits to the experimental data. (d) Representative field-free switching loops with various $J_{max}$ (in units of $10^{10}$ A/m$^2$). Switching percentages as functions of the applied $J_{max}$ for (e) pulse-width dependent and (f) temperature dependent measurements.

FIG1

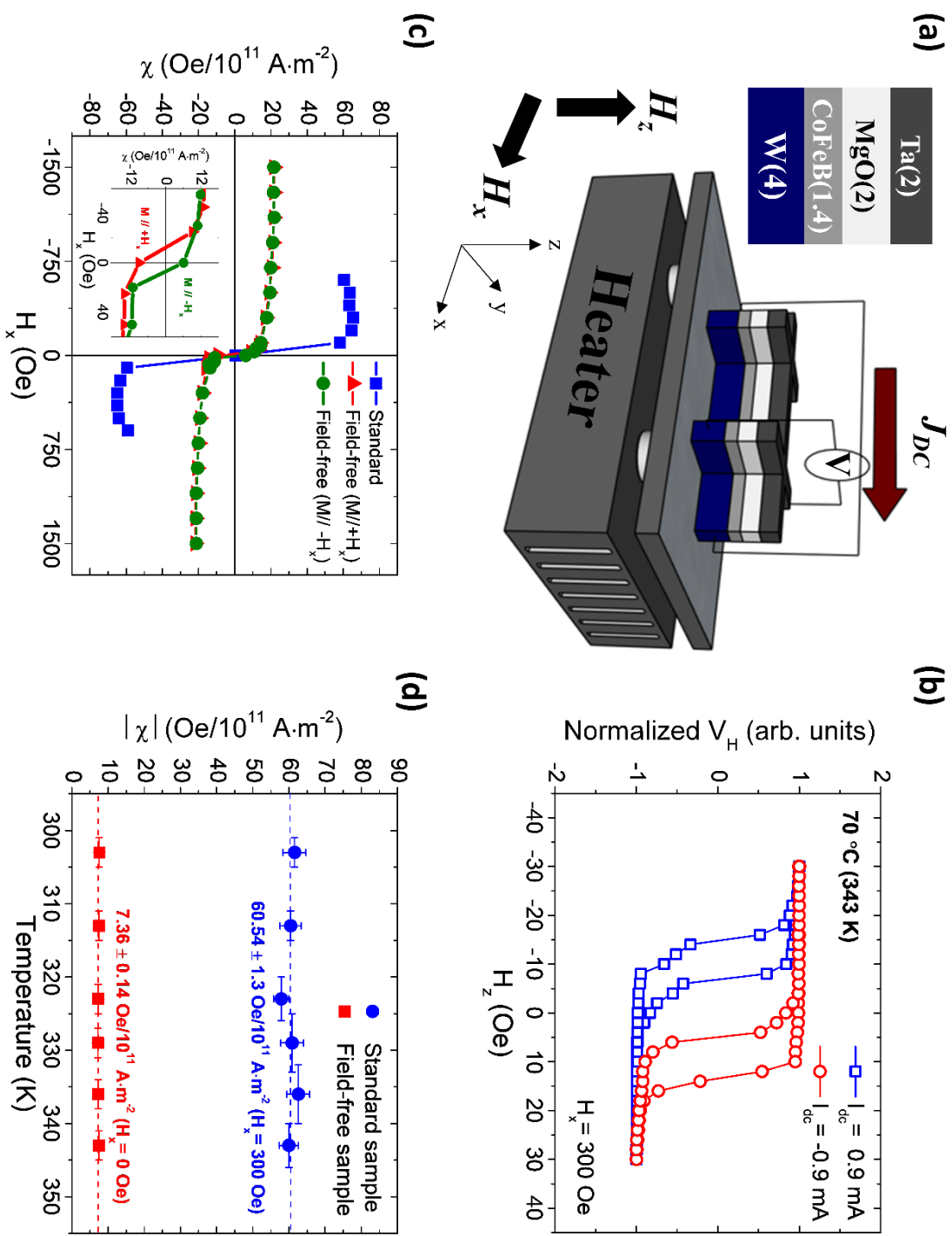

FIG2

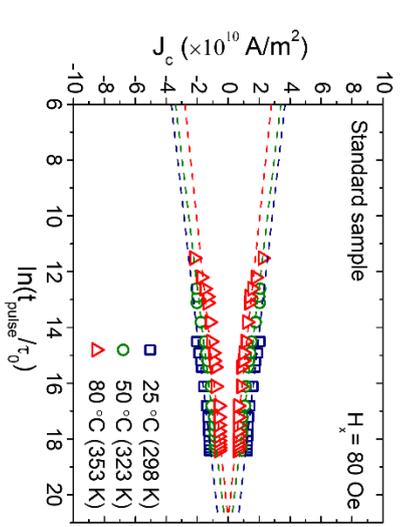
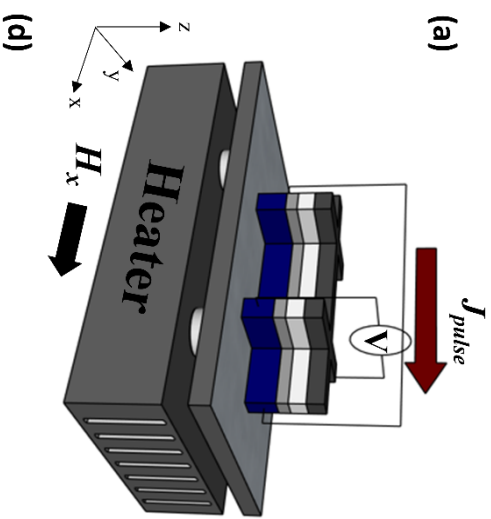

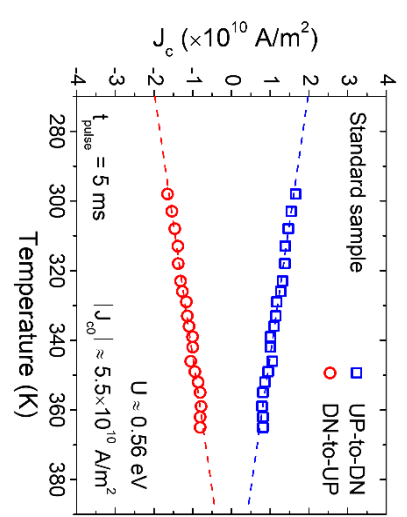
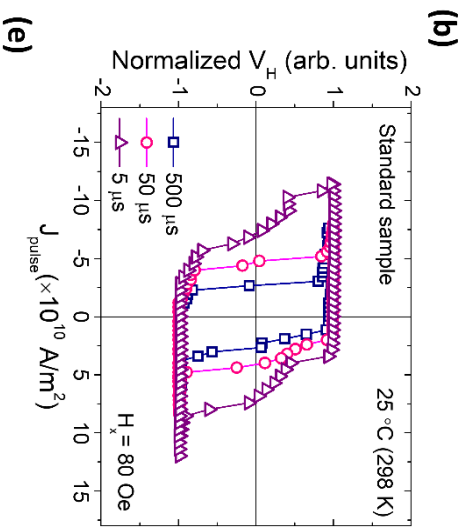

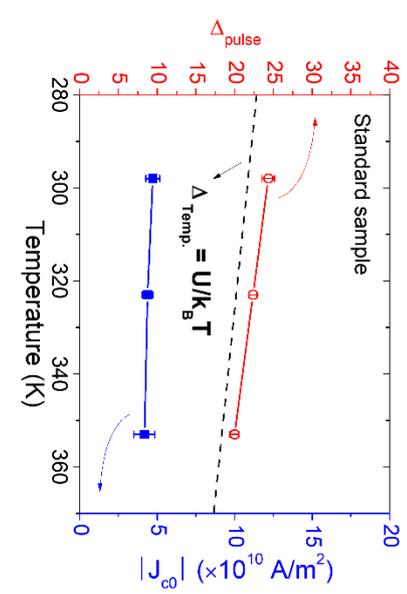
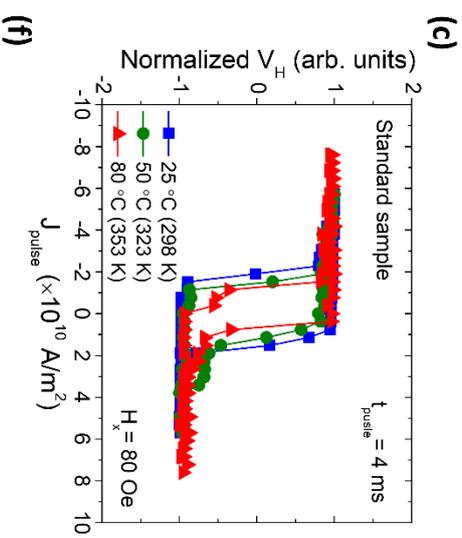

FIG3

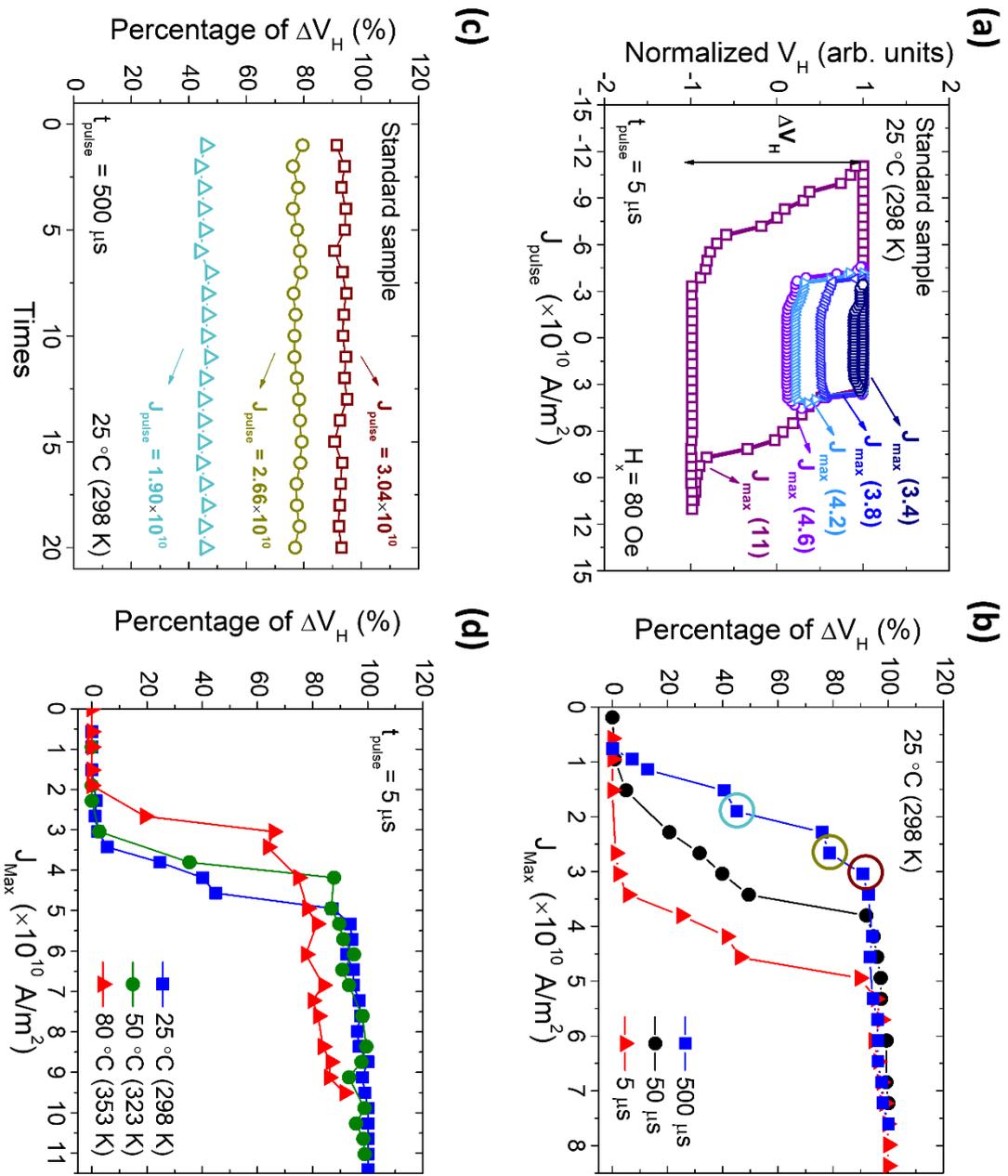

FIG4

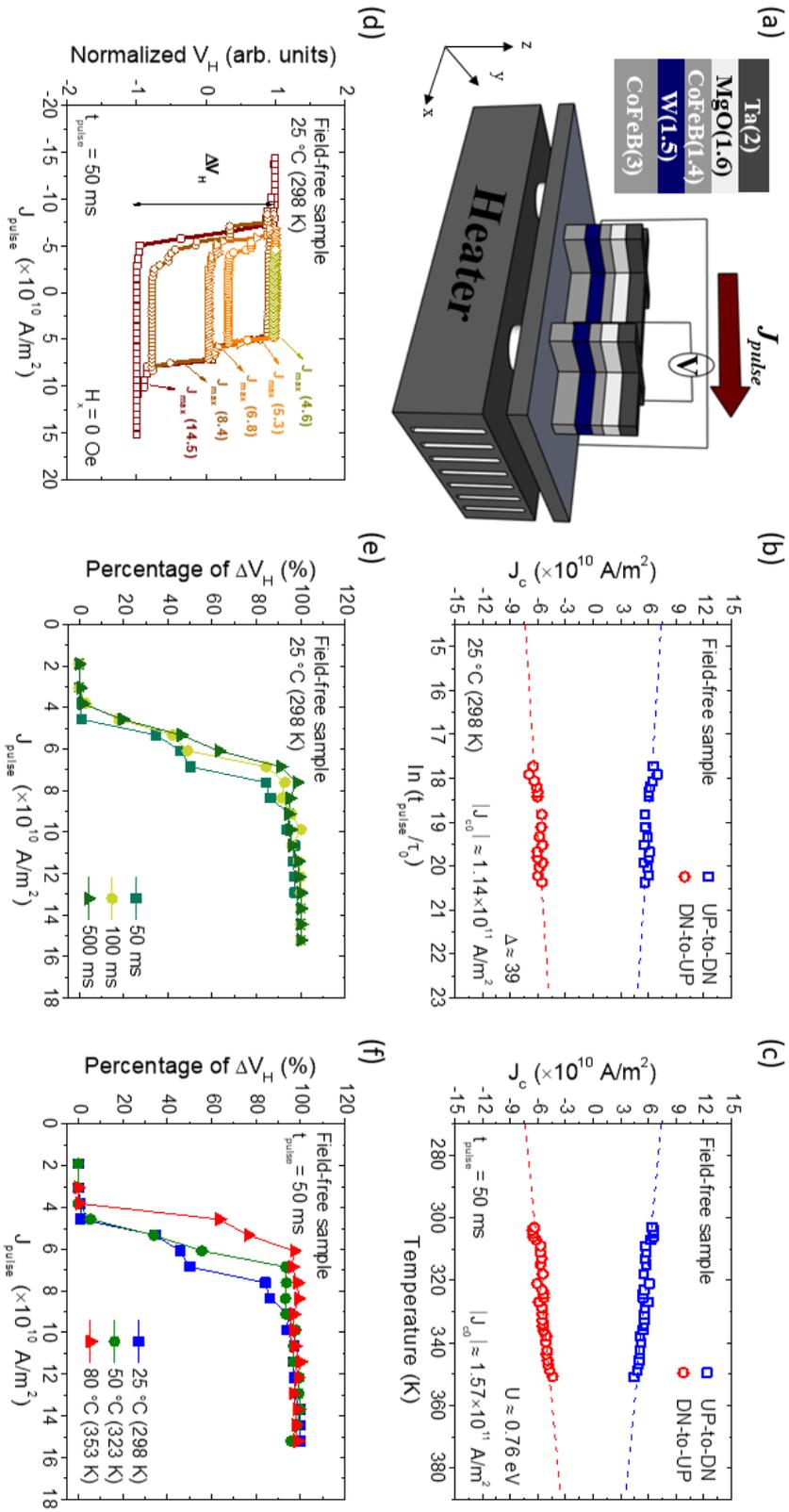